\documentclass[journal=achre4,manuscript=article,layout=twocolumn]{achemso}

\usepackage[T1]{fontenc}
\usepackage{graphicx}
\usepackage{amsmath}
\usepackage{amssymb}
\usepackage{tikz}
\usetikzlibrary{arrows.meta,positioning,calc}
\usepackage{xcolor}
\setkeys{acs}{articletitle=true,doi=true}
\AtBeginDocument{\singlespacing}

\title{On-Water Surface Catalysis: From Hydrogen Bonding to Charge-Transfer Activation}

\author{M. Alaraby Salem}
\affiliation{Dynamics of Condensed Matter and Center for Sustainable Systems Design, Chair of Theoretical Chemistry, Paderborn University, Warburger Stra\ss e 100, 33098 Paderborn, Germany}

\author{Thomas D. K{\"u}hne}
\email{tkuehne@cp2k.org}
\affiliation{Center for Advanced Systems Understanding (CASUS), Conrad-Schiedt-Stra{\ss}e 20, 02826 G{\"o}rlitz, Germany}
\alsoaffiliation{Helmholtz-Zentrum Dresden-Rossendorf, Bautzner Landstra{\ss}e 400, 01328 Dresden, Germany}
\alsoaffiliation{Institute of Artificial Intelligence, Technische Universit{\"a}t Dresden, Helmholtzstra{\ss}e 10, 01069 Dresden, Germany}

\keywords{on-water catalysis, aqueous interfaces, hydrogen bonding, proton transfer, charge transfer, ALMO-EDA, nuclear quantum effects, microdroplets}

\begin{document}

\begin{tocentry}
\centering
\includegraphics[width=3.0in]{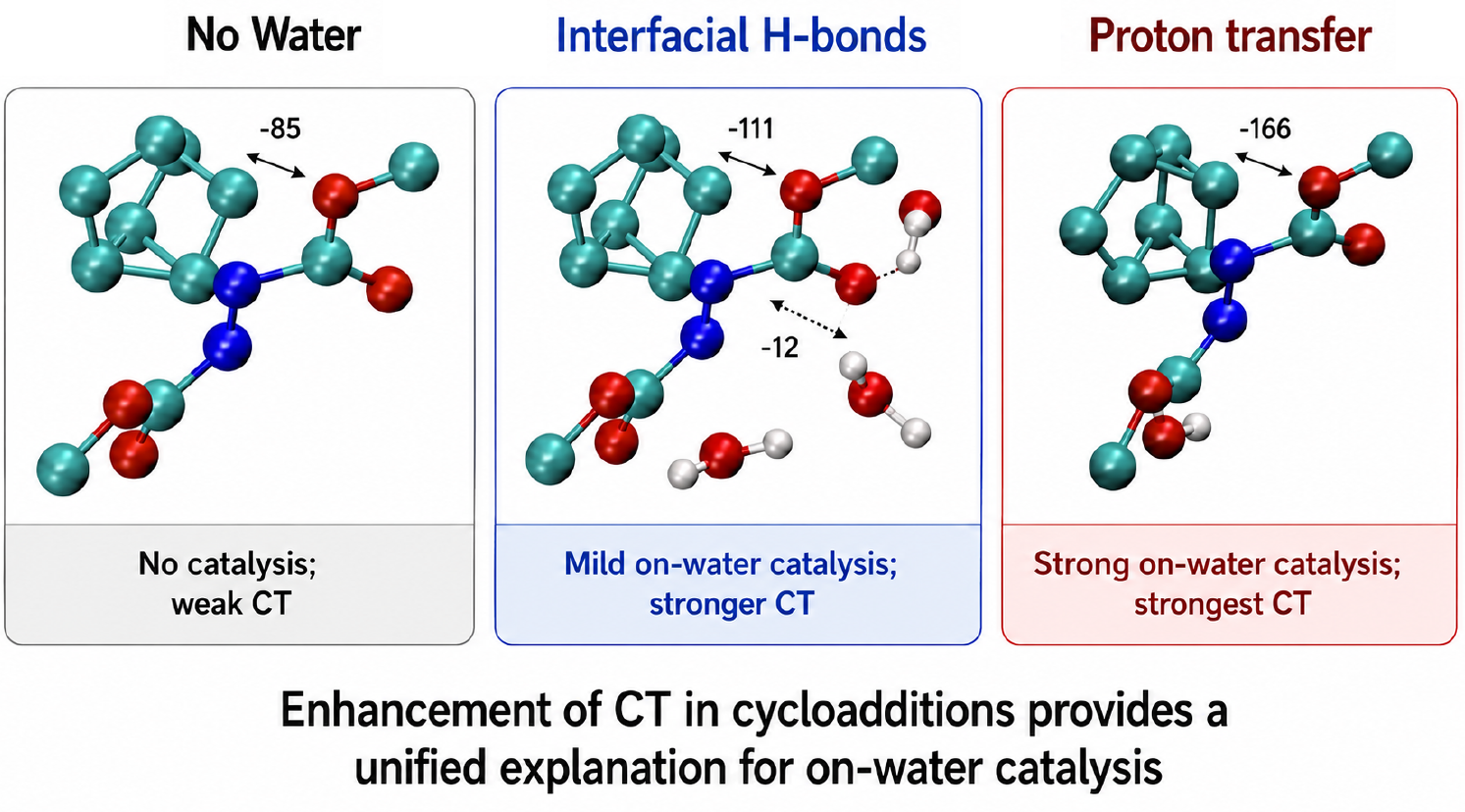}
\end{tocentry}

\begin{abstract}
Water is often introduced into organic synthesis as a solvent, a heat sink, or a benign alternative to hazardous media. A more intriguing phenomenon is on-water catalysis, where reactions between poorly soluble organic substrates proceed faster when the reactants are vigorously stirred in water. Since the seminal work of Sharpless and co-workers, two main mechanistic pictures have dominated the discussion. The first attributes the acceleration to hydrogen bonding from dangling OH groups at the water surface, which preferentially stabilize the transition state. The second emphasizes proton transfer at the interface, motivated by solvent kinetic isotope effects and analogies to acid catalysis.

This Account argues that both hydrogen bonding and proton transfer can enhance charge-transfer stabilization between the organic reactants. Hydrogen bonds from surface water polarize the reacting organic complex and enhance charge transfer between the organic partners. Activation by protonation can perturb the same donor--acceptor interaction more strongly, connecting the two pictures electronically without requiring identical elementary reaction pathways. We use energy decomposition analysis based on absolutely localized molecular orbitals, which can also be applied to condensed-phase systems. This analysis distinguishes direct interactions with water from the charge-transfer stabilization induced between the organic reactants.

This picture builds on a series of complementary studies. Simulations of the water/vapor interface established the presence of dangling OH groups capable of interacting with organic molecules. Finite-temperature simulations of on-water cycloadditions showed that a simple increase in transition-state hydrogen-bond count cannot explain their activation. Our first energy decomposition analysis then identified an indirect electronic effect in which hydration enhanced charge-transfer stabilization between the organic partners by approximately 30\%. A recent extension compares related dienophiles and a protonated limiting model. A complementary oxygen/sulfur substrate comparison links persistent hydration to different free-energy profiles and bond-formation pathways.

The resulting response metric quantifies the additional organic charge-transfer stabilization induced by controlled hydration. It provides a molecular hypothesis for why similar water contacts can have different catalytic consequences and suggests tests using matched substrate series, isotope substitution, and electronic analysis along the reaction coordinate. Activation ultimately depends on differential stabilization of reactant and transition-state ensembles, not on a single interaction component alone.

The unifying proposal is that interfacial water enhances charge-transfer stabilization between the organic reactants. Increasing the accessible reactive interface can amplify the contribution of this local mechanism without requiring exceptionally strong additional electric fields. This connects conventional on-water chemistry with selected microdroplet reactions while keeping adsorption, concentration, and intrinsic molecular activation distinct. Hydrogen bonds provide the contact. The substrate's electronic response determines what that contact can accomplish.
\end{abstract}

\section*{Key References}
\begin{itemize}
  \item Karhan, K.; Khaliullin, R. Z.; K{\"u}hne, T. D. On the Role of Interfacial Hydrogen Bonds in \textquotedblleft On-Water\textquotedblright{} Catalysis. \textit{J. Chem. Phys.} \textbf{2014}, \textit{141}, 22D528.\cite{karhan2014} Explicit finite-temperature simulations tested preferential transition-state hydrogen bonding and separated homogeneous hydration from interfacial effects.
  \item Salem, M. A.; K{\"u}hne, T. D. Insight from Energy Decomposition Analysis on a Hydrogen-Bond-Mediated Mechanism for On-Water Catalysis. \textit{Mol. Phys.} \textbf{2020}, \textit{118}, e1797920.\cite{salem2020} Electronic analysis identified water-enhanced charge transfer between the organic partners, beyond their direct interactions with water.
  \item Henao, A.; Gohar, Y.; Wilhelm, R.; K{\"u}hne, T. D. On the Role of Hydrogen Bond Strength and Charge Transfer in an On-Water Diels--Alder Reaction: Semiempirical and Free Energy Calculations. \textit{J. Comput. Chem.} \textbf{2026}, \textit{47}, e70467.\cite{henao2026} Comparing oxygen- and sulfur-containing dienophiles connected persistent hydration with different free-energy profiles and bond-formation pathways.
  \item K{\"u}hne, T. D.; Khaliullin, R. Z. Electronic Signature of the Instantaneous Asymmetry in the First Coordination Shell of Liquid Water. \textit{Nat. Commun.} \textbf{2013}, \textit{4}, 1450.\cite{kuhne2013} Electronic donor--acceptor asymmetry showed why similar water coordination geometries need not imply equal interaction strengths.
\end{itemize}

\section{Why On-Water Catalysis Still Needs a Mechanism}
Water is widely used in organic synthesis because it is abundant, inexpensive, nonflammable, and environmentally benign. Yet one of its most intriguing roles is in \emph{on-water} catalysis, where poorly soluble organic substrates react faster in vigorously stirred aqueous suspensions than in neat or organic media.\cite{narayan2005} The term emphasizes a heterogeneous reaction environment rather than conventional dissolution. An interfacial pathway is plausible, but the operational label alone does not locate every reactive event. Dissolved and organic-phase reactants may also contribute.

This interfacial nature makes the molecular origin of the catalytic effect difficult to establish. Poor solubility limits the applicability of conventional bulk-solvent models, while the dependence on stirring points to an important role for the water--organic interface. Different reactions also exhibit different catalytic enhancements and solvent kinetic isotope effects, which a unifying framework must connect to substrate-specific molecular responses. Two main explanations have dominated the discussion: stabilization of the transition state through hydrogen bonding from dangling OH groups at the water surface, and proton transfer from interfacial water to the reacting organic molecules.\cite{jung2007,beattie2010,butler2010,kitanosono2020}

This Account develops a unified framework that connects these two mechanistic pictures. Rather than treating hydrogen bonding and activation by protonation as competing explanations, we ask how both modify the same organic donor--acceptor interaction. Our published analysis and its recent extension\cite{salem2020,alaraby2026revision} provide the electronic foundation through absolutely localized molecular orbital energy decomposition analysis (ALMO-EDA). The resulting charge-transfer (CT) response measures the additional stabilization between organic reactants induced by hydration. Within this framework, the role of water extends beyond donating hydrogen bonds or transferring a proton. Interfacial water polarizes the organic reactants and can enhance charge-transfer stabilization between them.

On-water acceleration has been reported for a wide range of reactions, including cycloadditions, ene reactions, Claisen rearrangements, nucleophilic substitutions, and acid- or base-sensitive transformations.\cite{narayan2005,chanda2009,butler2010,kitanosono2020,li2021} The relative importance of hydrophobic assembly, interfacial adsorption, hydrogen bonding, proton transfer, and local electric fields varies across these reaction classes, suggesting that no single structural descriptor can account for all observations. This diversity motivates the need for a quantitative response metric. Rather than focusing on a specific interaction, such a metric should explain why closely related substrates exhibit different catalytic enhancements and solvent kinetic isotope effects.
\section{The Diels--Alder Reaction as the Test Case}
\begin{figure*}[t]
\centering
\begin{tabular}{cc}
\includegraphics[width=0.54\textwidth]{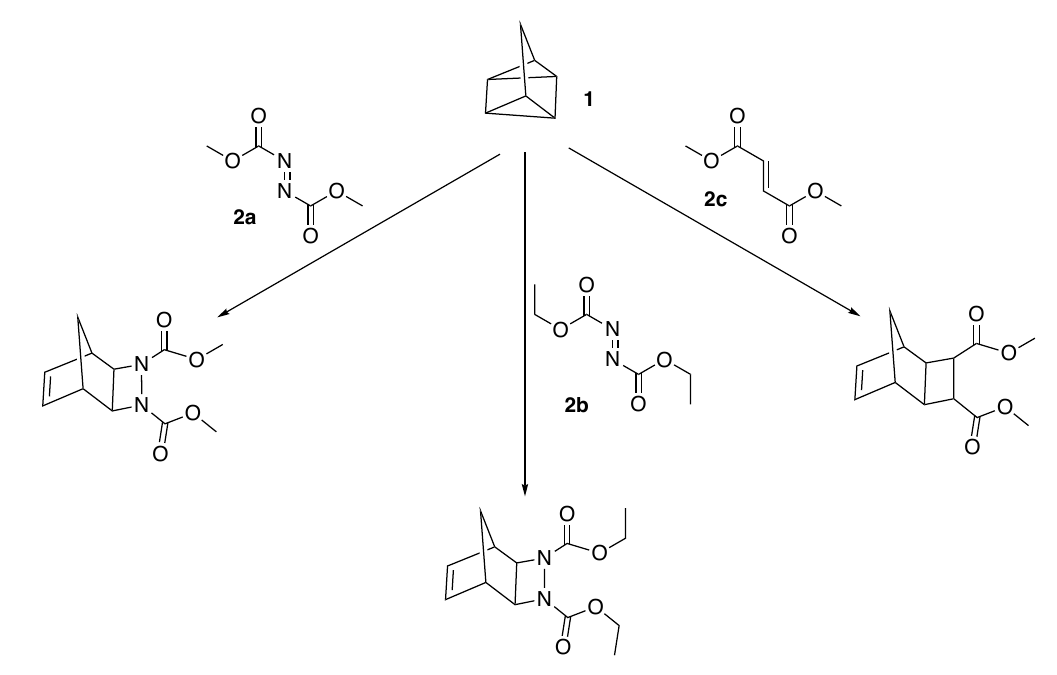} &
\includegraphics[width=0.36\textwidth]{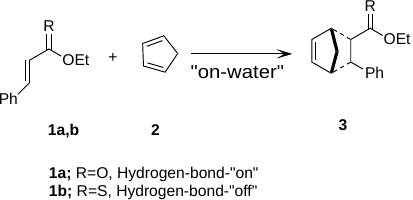}
\end{tabular}
\caption{Cycloaddition model systems used as the mechanistic thread of this Account. Left: quadricyclane (\textbf{1}) with azodicarboxylate and olefinic partners (\textbf{2a--c}). The hydration-response analysis uses \textbf{2a} and \textbf{2c}. Right: cyclopentadiene with ethyl cinnamate or thionocinnamate, used to compare hydration and bond formation at an aqueous interface. Left adapted from the authors' cycloaddition study.\cite{alaraby2026revision} Right adapted from Henao et al.,\cite{henao2026} licensed under CC BY 4.0.}
\label{fig:dielsmodels}
\end{figure*}

This Account uses Diels--Alder and related cycloaddition chemistry as the worked example because the field's central mechanistic dispute can be followed from observation to electronic response. These reactions connect the Sharpless observation, the Jung--Marcus dangling-OH model, the Beattie/Butler isotope-effect challenge, and our CT-response framework. Small changes to the organic partner alter hydrogen-bond acceptance, proton affinity, and electronic susceptibility.\cite{alaraby2026revision} The two families in Figure~\ref{fig:dielsmodels} provide complementary tests. Quadricyclane is not a conventional conjugated diene, so its cycloadditions are distinguished from the cyclopentadiene Diels--Alder reactions. Earlier simulations and free-energy calculations supply the condensed-phase context.\cite{henao2026,salem2020,karhan2014}

The appeal of these families is that the reaction coordinate is chemically transparent. Two new bonds form, and the transition-state region involves polarization and charge redistribution between organic partners. Water can influence the reaction in three distinguishable ways: by favoring productive interfacial encounters, by donating hydrogen bonds to acceptor groups, and by changing CT stabilization between the organic partners. These cycloadditions provide a platform for separating structural contact, proton motion, and CT stabilization.

\section{The Structural Premise: Dangling OH Groups Are Real}
The Jung--Marcus mechanism built on an intuitive structural feature of aqueous hydrophobic interfaces. Top-layer water molecules cannot complete a bulk-like tetrahedral hydrogen-bond network and therefore expose dangling OH groups toward the hydrophobic or vapor phase.\cite{jung2007} Surface-specific vibrational spectroscopy made this picture chemically compelling, and first-principles simulations supplied an atomistic complement.\cite{kuhne2011,kessler2015,bonn2015}

Our simulations sharpened the premise without turning it into a complete mechanism. First-principles molecular dynamics confirmed dangling OH bonds in the topmost layer and found no significant population of acceptor-only interfacial species.\cite{kuhne2011} Later path-integral and second-generation Car--Parrinello simulations resolved a layered motif with distinct orientational preferences.\cite{kessler2015} A dangling OH is a temporarily unaccepted donor bond, not a detached water molecule, and an organic adsorbate can reorganize this surface structure.

Our subsequent studies connected electronic hydrogen-bond asymmetry to vibrational frequencies and short-time water reorientation.\cite{kuhne2013,ojha2018,zysk2026} Simulations of surface-specific vibrational sum-frequency generation spectra further resolved distinct dynamics of free and hydrogen-bonded OH populations.\cite{ojha2019,ojha2021} These results support a picture in which an organic substrate encounters a distribution of contact geometries and electronic strengths rather than identical, static catalytic sites.

This matters because the CT-response picture still requires a chemically available interfacial perturbation. An outward-pointing OH exposes a partially positive hydrogen atom. That polarity already exists in an isolated water molecule. Orientation, polarization, and intermolecular electron redistribution modify its local environment.\cite{poli2020,leung2010} Three CT processes must be distinguished: redistribution within the water network, water--substrate CT, and CT between the organic reacting partners. The amplification discussed here concerns the third and need not imply net surface charging.

\begin{figure*}[t]
\centering
\includegraphics[width=0.98\textwidth]{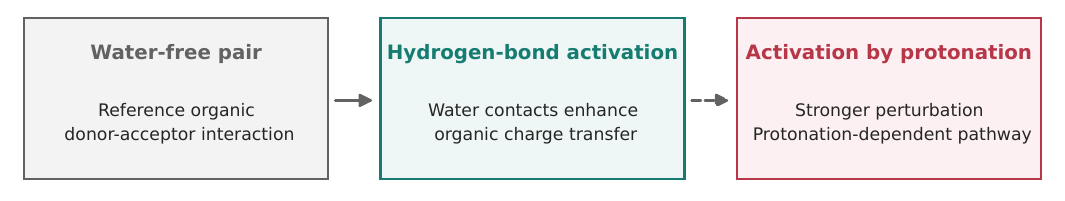}
\caption{
Conceptual connection between hydrogen-bond activation and activation by protonation. Both can enhance donor--acceptor stabilization within the organic reacting pair, relative to its water-free reference. The arrows connect electronic activation regimes rather than a calculated reaction path or a sequence of monotonically decreasing free-energy barriers. A change in protonation state additionally requires a defined proton source and thermodynamic reference.
}
\label{fig:unified}
\end{figure*}

\section{Testing the Jung--Marcus Picture}
Jung and Marcus translated the dangling-OH picture into a specific mechanism for the cycloaddition of quadricyclane with dialkyl azodicarboxylates.\cite{jung2007} In their model, interfacial water lowers the activation barrier because free OH groups form stronger and more numerous hydrogen bonds to the transition-state complex than to the reactants. This was an important conceptual advance because it linked the macroscopic need for heterogeneity to a molecular feature of the water surface.

The same model also made testable predictions. If a related substrate formed weaker hydrogen bonds to water, the on-water effect should diminish. Moreover, the rate acceleration should be tied primarily to preferential transition-state hydrogen bonding. These predictions became pressure points for the mechanism. Experiments and later analyses showed that related reactions can remain accelerated even when simple hydrogen-bond arguments suggested otherwise, and solvent isotope effects suggested that proton motion could not be ignored.\cite{beattie2010,butler2010,kitanosono2020}

Our 2014 simulation study directly addressed the condensed-phase assumptions behind the static cluster picture.\cite{karhan2014} Second-generation Car--Parrinello molecular dynamics\cite{kuhne2007,kuhne2014} with self-consistent-charge density-functional tight binding (SCC-DFTB)\cite{elstner1998} in CP2K\cite{wu2026water} enabled explicit finite-temperature sampling of the quadricyclane/dimethyl azodicarboxylate cycloaddition. Calculated barriers were approximately 72, 50, 57, and 59~kJ~mol$^{-1}$ for the isolated reaction, homogeneous aqueous solvation, and water/vapor and water/organic interfaces, respectively. Interfacial water remained important but did not outperform homogeneous hydration. For this reaction, water-induced barrier lowering is therefore not restricted to the interface. At the water/organic interface, the mean transition-state hydrogen-bond count increased by only 0.03 relative to the reactants. Static hydrogen-bond counting was not enough. The near-constant count directs attention to how existing water contacts affect the reacting pair, rather than to the recruitment of additional hydrogen bonds.

The complementary SCC-DFTB free-energy study, now published by Henao et al.,\cite{henao2026} compared cyclopentadiene reacting with ethyl cinnamate and ethyl thionocinnamate at an aqueous interface. Replacing the carbonyl oxygen by sulfur weakens water contacts, while also changing polarizability, intrinsic reactivity, and competing interactions. The comparison is therefore more informative than a literal hydrogen-bond-on/off switch because it relates hydration to the electronic and geometrical evolution of bond formation.

\begin{figure*}[t]
\centering
\includegraphics[width=0.95\textwidth]{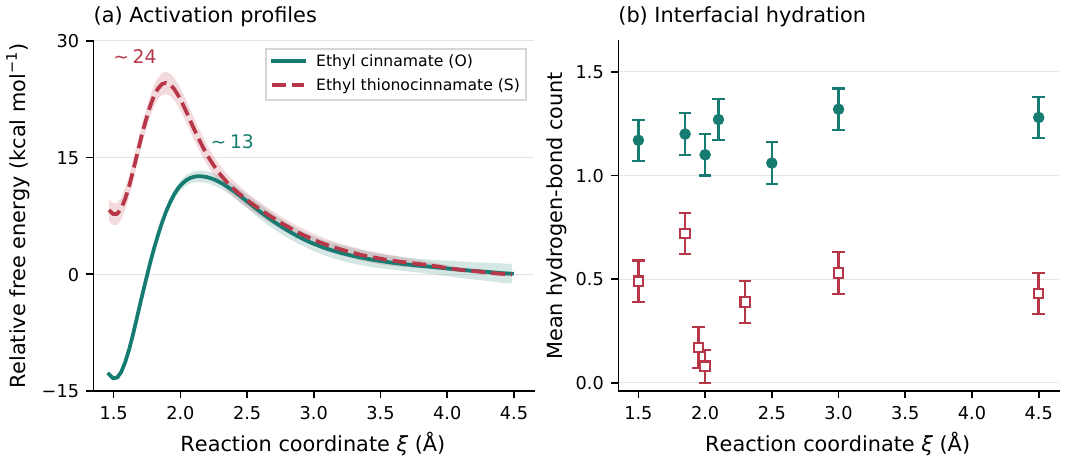}
\caption{Free energy and hydration along the same reaction coordinate for cyclopentadiene reacting with ethyl cinnamate (O, teal) or ethyl thionocinnamate (S, red) at an aqueous interface. (a) Relative free-energy profiles with the reported bootstrap uncertainty bands. (b) Mean water--dienophile hydrogen-bond counts with reported error bars. The coordinate $\xi$ is the average of the two forming carbon--carbon distances. Reaction proceeds from reactants near 4.5~\AA{} toward smaller $\xi$. Persistent oxygen hydration contrasts with loss of sulfur--water contacts near the barrier. The different substrates have different intrinsic reactivities, so their barrier difference is not an isolated water-induced catalytic effect. Redrawn without refitting from Figures 3 and 6 of Henao et al.,\cite{henao2026} licensed under CC BY 4.0.}
\label{fig:hbondonoff}
\end{figure*}

The results supply an important mechanistic connection. The oxygen-containing system has a barrier of approximately 13~kcal~mol$^{-1}$ near $\xi=2.1$~\AA{}, compared with 24~kcal~mol$^{-1}$ near $\xi=1.9$~\AA{} for the sulfur analogue. The carbonyl system maintains approximately 1.2 water hydrogen bonds along the pathway. Sulfur--water contacts instead fall from roughly 0.5 to almost zero near the barrier, accompanied by a strongly asynchronous pathway and a competing sulfur--carbon contact. The fractional contact counts are averages over sampled configurations, not numbers of hydrogen bonds in a single structure. The reported transition-state difference between the forming carbon--carbon distances changes from $0.02\pm0.20$ to $0.95\pm0.07$~\AA{} upon oxygen/sulfur substitution. For the oxygen-containing system, the near-zero difference means that both bonds are similarly developed at the transition state on average. In the sulfur analogue, one forming bond is almost 1~\AA{} shorter and therefore substantially more developed than the other. The transient contact between sulfur and a cyclopentadiene carbon competes with formation of the second carbon--carbon bond.\cite{henao2026} Productive hydration and bond-formation geometry must thus be considered together. This points toward the ALMO-EDA question: how does water sustain electronic redistribution in the reacting pair without an increase in hydrogen-bond count?

\section{Proton Transfer and Isotope Effects}
Beattie, McErlean, and Phippen did not introduce proton transfer chemistry to aqueous interfaces, but they provided one of the clearest on-water formulations of a proton-transfer mechanism.\cite{beattie2010} In their picture, interfacial water activates organic substrates through proton transfer, with hydroxide stabilization at the oil--water interface driving the equilibrium. This model naturally accommodates solvent kinetic isotope effects and the observation that many on-water reactions are also acid-catalyzed.\cite{butler2010,kitanosono2020} It also emphasizes a point sometimes blurred in purely structural models. The interface differs from bulk water not only in geometry but also in acid--base and electrostatic character.

For the field, this argument exposed what a static hydrogen-bond picture could not easily explain. The same literature emphasized on-water acceleration of supposedly weakly hydrogen-bonding dienophiles. Yet a solvent kinetic isotope effect (KIE), such as $k_{\mathrm{H_2O}}/k_{\mathrm{D_2O}}$ under matched conditions, is not unique evidence of proton transfer. A value above unity means that the reaction is faster in H$_2$O than in D$_2$O. Both hydrogen bonding and proton-transfer chemistry involve isotope-sensitive nuclear motion. The challenge is to connect their electronic consequences while identifying the regime relevant to each reaction.

This is where nuclear quantum effects enter the story. Zero-point motion and isotope-dependent hydrogen-bond fluctuations alter the nuclear distributions sampled at an interface.\cite{kessler2015,habershon2009} Isotope substitution can also change proton availability and barrier-crossing dynamics. A measured KIE therefore constrains a mechanism but does not assign a unique position on a polarization continuum. Conversely, a small KIE does not imply that interfacial water is unimportant.

The proposed electronic connection arises because hydrogen-bond donation and proton sharing can perturb the same substrate orbitals, whereas a change in protonation may introduce a distinct reaction pathway. Isotope-resolved free energies and proton-activity controls can identify when that additional pathway becomes accessible. Equilibrium path-integral sampling captures statistical isotope effects, but is not by itself a calculation of a real-time rate.

\section{Energy Decomposition: From Hydrogen Bonds to Electronic Response}
The central weakness of a hydrogen-bond-counting picture is that it treats water--substrate interactions as the catalytic object. Yet a reaction barrier is controlled by the relative stabilization of the transition-state region compared with the reactants. The catalytic object is therefore the entire reacting complex, including how water changes electron sharing and polarization within it.

ALMO-EDA, as implemented in the CP2K program package,\cite{cp2k2020,cp2k2026} provides a useful way to make this statement quantitative.\cite{khaliullin2007,khaliullin2009,khaliullin2013pccp,elgabarty2015,kuhne2013,salem2020} At a fixed geometry and for specified fragments, the interaction energy is separated into frozen, polarization, and CT contributions:
\begin{equation}
\Delta E_{\mathrm{int}}=\Delta E_{\mathrm{frz}}+\Delta E_{\mathrm{pol}}+\Delta E_{\mathrm{CT}}.
\end{equation}
The frozen term includes electrostatic and Pauli contributions from unrelaxed fragment densities. Polarization permits intrafragment relaxation while preventing interfragment delocalization. CT measures the additional stabilization when that constraint is lifted. Water--substrate CT is therefore neither the full hydrogen-bond energy nor the polarization energy. For on-water catalysis, the key advantage is that direct interactions with water can be distinguished from the electronic response between the two organic reactants.

Here, charge transfer denotes electronic delocalization between the organic reactants, not an electron-transfer reaction. Its enhancement by protonation does not, by itself, establish a proton-coupled electron-transfer mechanism.

Assigned transferred charges and CT stabilization energies are different quantities. ALMO-based charge analysis can assign smaller transferred electron populations than conventional population analyses without implying an energetically negligible interaction.\cite{khaliullin2007} Elgabarty et al. connected ALMO-based CT descriptors in liquid water to proton magnetic shielding, providing an indirect spectroscopic probe of hydrogen-bond covalency.\cite{elgabarty2015} The comparisons below use a consistent ALMO definition. Phenomenological conclusions should be assessed against free energies, spectra, or kinetics rather than the magnitude of one charge partition.

The 2020 ALMO-EDA analysis is the key proof of principle beneath the later CT-response metric.\cite{salem2020} We revisited the Jung--Marcus configurations for the same quadricyclane/dimethyl azodicarboxylate cycloaddition shown in Figure~\ref{fig:dielsmodels}. The hydrogen bonds between water and the dienophile were not markedly stronger in the transition state than in the reactant configuration. Depending on the energy component, the reactant-side interaction could appear equally strong or more favorable. The direct water--dienophile CT interaction remained essentially constant at about $-12$~kJ~mol$^{-1}$. Instead, water increased organic reactant--reactant CT stabilization in the transition-state geometry from $-84.7$ to $-110.9$~kJ~mol$^{-1}$, an enhancement of roughly 30\%. More negative CT energies denote stronger stabilization through electronic delocalization between the organic reactants. The 30\% refers to this energy component, not to a 30\% decrease in the activation barrier.

This result reframes the role of water. Interfacial water polarizes the dienophile and makes the electronic interaction between the organic partners more favorable. The hydrogen bond is the handle. Enhanced reactant--reactant CT is the electronic response. The later comparative analysis extends this proof of principle by asking how strongly different reacting pairs respond to hydration. These findings suggest a common electronic basis for hydrogen-bond activation and activation by protonation. Sharpless established the on-water phenomenon, and Jung and Marcus identified the chemically active dangling-OH motif. Isotope-effect and proton-transfer arguments brought proton motion into the discussion. ALMO-EDA connects these observations in one electronic response picture.

\section{A CT-Response Metric for Interfacial Polarization}
The central development of this Account is to turn the qualitative reframing into a comparative CT-response metric, a compact measure of how strongly a reacting pair responds electronically to water.\cite{alaraby2026revision} A particularly important case is the comparison between azodicarboxylate partners and olefinic analogues in quadricyclane cycloadditions. Earlier hydrogen-bond arguments suggested that replacing nitrogen atoms should weaken water activation. Experimentally, however, related olefinic systems can still show on-water acceleration.\cite{beattie2010}

Our recent ALMO-EDA extension supplies an electronic explanation for this apparent contradiction.\cite{alaraby2026revision} In the transition-state-derived cluster of the olefinic analogue, water hydrogen bonds to carbonyl oxygen atoms rather than to the atoms introduced by replacing nitrogen. Productive hydration is therefore retained. More importantly, hydrogen-bond strength alone does not specify how strongly that hydration changes donor--acceptor stabilization between the organic partners.

To separate intrinsic coupling from its hydration response, we define
\begin{equation}
S_{\mathrm{org}}(N_{\mathrm w})=
E_{\mathrm{CT}}^{\mathrm{org}}(0)-E_{\mathrm{CT}}^{\mathrm{org}}(N_{\mathrm w}),
\label{eq:ctresponse}
\end{equation}
where $N_{\mathrm w}$ is the number of retained waters. Positive $S_{\mathrm{org}}$ denotes enhanced organic CT stabilization relative to the same water-free geometry. Waters are removed from a fixed transition-state-derived cluster, providing a controlled electronic perturbation. Holding the nuclei fixed separates this electronic response from molecular rearrangement. For \textbf{2a}, the organic CT term changes from $-84.7$ to $-110.9$~kJ~mol$^{-1}$ between zero and three waters. For \textbf{2c}, it changes from $-128.3$ to $-146.2$~kJ~mol$^{-1}$. Thus, the more strongly coupled dry pair \textbf{2c} has a smaller hydration response of 17.9 rather than 26.2~kJ~mol$^{-1}$. 

\begin{figure}[!t]
\centering
\includegraphics[width=\columnwidth]{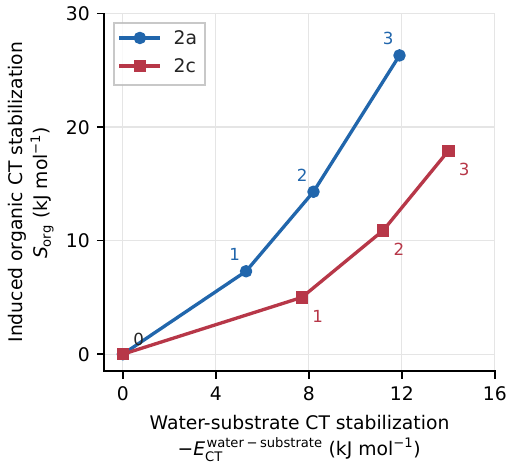}
\caption{Electronic response to controlled hydration. The additional charge-transfer stabilization between organic partners, relative to each dry geometry, is plotted against the magnitude of water--substrate charge-transfer stabilization. Labels indicate zero to three retained water molecules. The dry references coincide at the origin because the additional stabilization is zero by definition, despite their different absolute CT energies. Lines connect the fixed-geometry results, not a fitted kinetic relationship. The stronger dry coupling of \textbf{2c} does not imply a larger hydration response. Values are from Table S3 of our cycloaddition manuscript.\cite{alaraby2026revision}}
\label{fig:ctresponse}
\end{figure}

Figure~\ref{fig:ctresponse} relates this response to the magnitude of water--substrate CT. For the three-water clusters, \textbf{2c} has the larger direct water--substrate CT magnitude, yet \textbf{2a} gains more organic CT stabilization. Thus, stronger direct CT with water does not necessarily produce a larger electronic response within the reacting pair. The slopes characterize these hydration series, not a universal conversion efficiency from polarization to reaction rate.

The same analysis provides a natural bridge to proton transfer. A protonated transition-state model for \textbf{2a} reaches an organic CT stabilization of approximately $-166$~kJ~mol$^{-1}$, nearly twice the magnitude of the dry value.\cite{alaraby2026revision} The same reacting pair therefore responds in the same stabilizing direction to hydration and to protonation, with a much larger response in the latter case. This is the electronic connection represented in Figure~\ref{fig:unified}. Hydrogen-bond activation and activation by protonation can share a common donor--acceptor response even when their elementary pathways differ.

This is the point at which the hydrogen-bond and proton-transfer literatures become mutually informative. Jung and Marcus identified the structural motif that makes interfacial water chemically available. Isotope-effect and acid-catalysis arguments brought proton motion into the discussion. ALMO-EDA connects these views electronically. Water contacts perturb the organic donor--acceptor interaction, and stronger proton sharing or protonation can amplify that response.

The quantitative boundary is straightforward. These cluster CT energies are interaction components, not activation free energies. Their kinetic significance requires comparing reactant and activated ensembles with consistent fragmentation, including polarization, repulsion, deformation, and entropy. The protonated model additionally requires a proton source, countercharge, and proton-activity reference. Its stabilization alone does not establish favorable proton transfer from neutral water. These requirements define how to test the proposed response metric against catalytic and isotope effects.

The framework also has boundaries. It does not deny that hydrophobic assembly, local concentration, or surface adsorption can dominate measured rates in some systems. The role of interfacial polarization and CT in activation depends on the reactants and the reaction pathway. To assess this role, the framework distinguishes two questions. The first is kinetic access: how many reactive complexes reach the interface and adopt productive geometries? The second is electronic activation: once a complex is at the interface, how does water change the transition-state energy? Interfacial polarization and CT-response address the second question. They should therefore be combined with, rather than substituted for, careful kinetic models of adsorption, phase transfer, and interfacial area.

\section{What This View Explains}
The interfacial-polarization picture brings together four features that are difficult to explain in isolation.

First, it preserves the core intuition of Jung and Marcus that dangling OH groups at aqueous hydrophobic interfaces are chemically important. Surface water can donate hydrogen bonds into the organic phase, and the orientational structure of the topmost interfacial layer provides the required geometry.\cite{kuhne2011,kessler2015,jung2007}

Second, it explains why counting hydrogen bonds is insufficient. A hydrogen bond can be structurally similar in reactant and transition-state configurations while still having a different catalytic consequence if it changes the electronic coupling between the organic reactants.\cite{henao2026,salem2020}

Third, it incorporates the isotope-based proton-transfer objection without making every on-water reaction a proton-transfer reaction. KIEs constrain proton-sensitive contributions to activation, while the electronic response links hydrogen-bonded and protonated regimes.\cite{beattie2010,butler2010,kitanosono2020}

Fourth, it shifts attention from a universal surface property to a reaction-specific response function. The same interfacial water motif can have different consequences depending on the acceptor ability, polarizability, proton affinity, and frontier-orbital structure of the organic reactants. This suggests evaluating substrates not only by their ability to hydrogen bond to water but also by their ability to translate that interaction into productive organic CT stabilization.

\section{Predictions and Tests}
The interfacial-polarization picture is useful only if it makes predictions that could fail. Several tests are immediate.

First, substrate series should separate hydrogen-bond acceptance from electronic response as far as chemistry permits. Dienophiles with similar carbonyl contact motifs but different acceptor levels or polarizabilities should show different hydration responses. We hypothesize that the differential organic CT response between reactant and activated ensembles will explain intrinsic barrier changes better than contact counts alone. Systematic failure after controlling adsorption and reference states would identify the limits of CT as a descriptor.

Second, isotope effects should be measured across matched substrate families with independently characterized proton affinities. Temperature and proton-activity dependence can help distinguish hydration-induced electronic activation from an accessible protonation-dependent pathway. The test is not whether all strongly accelerated reactions have large KIEs, but whether the proposed proton-sensitive contribution is consistent with isotope-resolved free energies and kinetics.

Third, interfacial area and interfacial structure should be controlled independently. Microfluidic platforms, adsorbed water in nanoporous materials, and well-characterized oil--water interfaces can separate area, adsorption, and intrinsic electronic effects.\cite{mellouli2012,guo2016,nguyen2021} At comparable surface composition and local reactivity, greater accessible area should increase total interfacial turnover without changing the molecular hydration response. Area-normalized rates, adsorption measurements, and controlled mixing provide a direct test.

Fourth, simulations should report more than activation barriers. Free-energy profiles remain essential, but an on-water mechanism should also report interfacial location, hydrogen-bond motifs, polarization components, proton-sharing coordinates, and fragment-resolved CT terms. This combination would allow the field to compare reactions mechanistically rather than only kinetically.

\section{Other Mechanistic Ideas}
On-water catalysis has also been discussed in terms of hydrophobic assembly, cohesive energy density, local concentration, Eley--Rideal surface kinetics, electric fields, and mass-transfer effects.\cite{breslow1991,engberts2001,chanda2009,guo2016,nguyen2021} These ideas are not mutually exclusive with the electronic picture. In fact, they operate at different levels.

Hydrophobic effects and interfacial adsorption can increase encounter probability and orient reactants. Controlled-interfacial-area experiments and nanoporous-water systems have shown that kinetics can depend sensitively on how reactants populate the interface.\cite{guo2016,nguyen2021} Such effects determine whether the reacting complex is placed in the right interfacial environment. The ALMO-EDA picture then asks what water does electronically once the complex is there.

Likewise, local electrostatics is part of the molecular activation picture. Hydrogen bonding combines electrostatic, polarization, and short-range orbital interactions. It is not reducible to a field magnitude alone. A local field can bias electron distribution and proton position, but a catalytic effect requires preferential stabilization of the activated ensemble relative to the reactants. These local interactions are already present in electronic-structure calculations with explicit water.

\section{Practical Opportunities for On-Water Catalysis}
The CT-response view makes on-water catalysis more actionable. The practical question is whether a substrate pair can be placed at an aqueous interface that activates its donor--acceptor interaction productively. Stirred emulsions, microfluidic droplets, flow reactors, and supported aqueous films offer control over contact area. Substrate design can expose acceptor groups that couple hydration to bond formation while avoiding competing contacts, hydrolysis, or unproductive protonation.

Surfactant-monolayer-assisted interfacial synthesis (SMAIS) offers a concrete connection to materials synthesis. Feng and co-workers use surfactant monolayers to preorganize monomers and direct the growth of crystalline two-dimensional polymers on water.\cite{liu2019smais} Seki et al. used surface-specific vibrational spectroscopy to identify positively charged aniline-derived intermediates that accumulate and become ordered beneath negatively charged surfactant headgroups, promoting crystalline polyaniline films.\cite{seki2021smais} This provides experimental support for electrostatic organization as a central contribution to SMAIS. Feng and co-workers also discuss hydrogen-bond stabilization of activated complexes as a possible contribution to enhanced surface reactivity.\cite{ni2024smais}

The O-SMAIS variant extends this strategy to organic--aqueous subphases. Yang et al. demonstrated that an anionic perfluorosurfactant monolayer on a 1:1 dimethylacetamide/water mixture by volume accumulates Cu$^{+}$ ions, creating a catalyst-rich interface. Coordination of terminal alkynes to these ions then directs monomer assembly and Glaser coupling into diyne-linked two-dimensional polymer crystals.\cite{yang2025osmais} Both charge-mediated accumulation and the structure of the monolayer contribute to the synthetic outcome.

Ionic surfactant charges, compensated by counterions and subject to screening, are distinct from the partial charges exposed by oriented water molecules. Neither should be equated with intermolecular CT. Electrostatic accumulation can be explained without invoking hydration-induced CT enhancement. Our framework immediately suggests testing whether surface charge additionally modifies hydration, protonation, and CT stabilization within the reacting complexes. Such electronic activation would complement, rather than replace, electrostatic accumulation and coordination-controlled assembly. Stronger adsorption alone need not lower the intrinsic barrier, since it can also stabilize the reactants. Comparing activation free energies and ALMO-EDA components along the reaction coordinate, at controlled interfacial populations and catalyst speciation, would test whether the environment preferentially stabilizes the transition state.

Reactions with poor bulk solubility, polarizable activated complexes, and products that separate from water are attractive candidates. Cycloadditions remain a useful development platform, with extensions to other reaction classes requiring matched tests. The distinction between encounter and activation matters for scale-up. Area and mixing control access, substrate and interface design control local electronic response, and proton activity controls the availability of protonation-dependent pathways.

Connecting those experimental variables to electronic response would turn on-water catalysis from a retrospective label into a design principle. The aim is to identify reactions where water-rich interfaces provide useful activation together with simpler separation or reduced organic-solvent demand.

\section{Microdroplets as an Extreme On-Water Limit}
Microdroplets offer a direct test of the interfacial interpretation. Bain, Sathyamoorthi, and Zare reported a further acceleration of approximately two orders of magnitude for the quadricyclane/diethyl azodicarboxylate cycloaddition relative to the earlier on-water experiment, while retaining similar solvent trends.\cite{bain2017} We propose that such acceleration can reflect greater access to the same water-induced electronic activation, rather than a distinct molecular mechanism.

The distinction between local activation and total turnover is central. If $j_{\mathrm{int}}$ is the reactive flux per unit accessible interface, the interfacial contribution to product formation per unit volume is
\begin{equation}
\frac{R_{\mathrm{int}}}{V}=\frac{A_{\mathrm{react}}}{V}\,j_{\mathrm{int}}.
\label{eq:area}
\end{equation}
For spherical droplets of radius $r$, the geometric surface-to-volume ratio is $3/r$. For a fixed total liquid volume, dividing it into droplets of one-tenth the radius gives ten times more surface area. If the reactive fraction of that surface and $j_{\mathrm{int}}$ remain unchanged, interfacial turnover per volume also rises tenfold without changing the local barrier. The relevant area is the water--organic contact sampled by reactants, not automatically the entire external droplet surface.

In this sense, microdroplets can be an area-amplified realization of on-water catalysis. Exceptionally strong additional electric fields are not required by this explanation. Establishing quantitative sufficiency for the reported enhancement requires measured reactive area, interfacial populations, and local rates. Evaporation and concentration changes must be separated. For a different cyclopentadiene/acrylonitrile reaction, combined calculations and experiments have identified confinement and evaporative enrichment, rather than field catalysis, as the principal sources of microdroplet acceleration.\cite{gong2024} That result illustrates the distinction between enhanced overall kinetics and a reduced intrinsic barrier.

The hypothesis predicts that, after controlling composition and adsorption, area-normalized rates and local electronic responses should connect conventional on-water and microdroplet experiments. It neither requires a universal microdroplet mechanism nor addresses unrelated redox chemistry. Its core claim is that more opportunities for the same productive water contact can produce faster overall chemistry.

\section{Outlook}
The central proposal of this Account is that hydrogen-bond activation and activation by protonation can be understood through a common electronic response in which water modifies donor--acceptor interactions within the organic reacting pair. This connects the structural insight of dangling-OH models to the electronic consequences of hydration and proton motion.

The next step is to combine finite-temperature sampling with complete energy decomposition along the reaction coordinate. Matched substrate series, isotope-resolved free energies, and controlled adsorption and interfacial-area measurements would test whether the response metric captures the differences that determine catalytic efficiency. This progression links the published electronic mechanism to a predictive description rather than treating a static CT energy as a rate constant.

The same picture provides a molecular basis for connecting conventional on-water chemistry with microdroplets. Greater accessible reactive area can amplify an existing local activation mechanism. Geometrical amplification and local electronic activation are complementary parts of this explanation.

Hydrogen bonds provide the first contact, and protonation can strengthen the perturbation. CT between the organic reactants is the electronic thread connecting them. Understanding how substrates translate water contacts into productive electronic response offers a route to designing aqueous interfaces for synthesis.

\section*{Author Information}
\textbf{Corresponding Author}\\
Thomas D. K{\"u}hne; Email: tkuehne@cp2k.org

\section*{Biographies}
M. Alaraby Salem is a theoretical chemist whose work focuses on energy decomposition analysis, hydrogen bonding, and electronic mechanisms at aqueous interfaces. His research has used ALMO-EDA to dissect how interfacial water modifies charge transfer in on-water catalysis.

Thomas D. K{\"u}hne is a theoretical chemist and scientific director at CASUS. His research develops and applies electronic-structure, molecular-dynamics, and machine-learning methods to hydrogen-bonded systems, aqueous interfaces, and chemical reactivity in complex environments.

\section*{Notes}
The authors declare no competing financial interest.

\section*{Acknowledgments}
The authors gratefully acknowledge Kristof Karhan, Rustam Z. Khaliullin, Jan Kessler, Hossam Elgabarty, Thomas Spura, Pouya Partovi-Azar, and Ali A. Hassanali for collaborations that shaped the mechanistic view summarized here.
Part of the research was funded by the Deutsche Forschungsgemeinschaft (DFG, German Research Foundation), project number 417590517/CRC1415.

\bibliography{account_references_v2,arabisch_revision_references}

\end{document}